\documentstyle[12pt]{article}

\catcode`\@=11 \@addtoreset{equation}{section}\catcode`\@=12
\newcommand{\be}{\begin{equation}}
\newcommand{\ee}{\end{equation}}
\newcommand{\ba}{\begin{eqnarray}}
\newcommand{\ea}{\end{eqnarray}}

\begin{document}
\thispagestyle{empty}

\begin{center}
               RUSSIAN GRAVITATIONAL SOCIETY\\
               INSTITUTE OF METROLOGICAL SERVICE \\
               CENTER OF GRAVITATION AND FUNDAMENTAL METROLOGY\\

\end{center}
\vskip 4ex
\begin{flushright}
                                         RGS-CSVR-002/96
                                         \\ hep-th/9603xxx

\end{flushright}
\vskip 15mm

\begin{center}
{\large\bf  Multidimensional Extremal Dilatonic Black
Holes in String-like Model with Cosmological Term}

\vskip 5mm
{\bf V. D. Ivashchuk and V. N. Melnikov }\\
\vskip 5mm
     {\em Center for Gravitation and Fundamental Metrology,
     VNIIMS, 3-1 M. Ulyanovoy St., Moscow, 117313, Russia}\\
     e-mail: mel@cvsi.rc.ac.ru \\
\end{center}
\vskip 10mm

ABSTRACT

A string-like model with the "cosmological constant" $\Lambda$
is considered. The Maki-Shiraishi multi-black-hole solution
\cite{MS1} is generalized to space-times with a
Ricci-flat internal space. For $\Lambda = 0$
the obtained solution in the one-black-hole case is shown
to coincide with the extreme limit of the charged dilatonic black hole
solution \cite{BI,BM}. The Hawking temperature $T_H$ for the
solution \cite{BI,BM} is presented and its extreme limit is considered.
For the string value of dilatonic coupling the temperature $T_H$
does not depend upon the internal space dimension.

\vskip 10mm

PACS numbers: 04.20, 04.40.  \\
Key words: Dilatonic Black Holes, Strings, Hawking Temperature \\

\vskip 30mm

\centerline{Moscow 1996}
\pagebreak

\setcounter{page}{1}

\pagebreak

%%%%%%%%%%%%%%%%%%%%%%%%%%%%%%%%%%%%%%%%%%%%%%%%%%%%%%%%%%%%%%%%%%
\section{Introduction}
\setcounter{equation}{0}
%%%%%%%%%%%%%%%%%%%%%%%%%%%%%%%%%%%%%%%%%%%%%%%%%%%%%%%%%%%%%%%%%%

In this paper we consider the model described by the
action
\begin{equation} %1.1
S = \int d^{D}x \sqrt{|g|} \{ \frac{1}{2 \kappa^{2}}
\Big[ {R}[g] -  2 \Lambda \exp(- 2 \lambda \varphi)
- \partial_{M} \varphi  \partial_{N} \varphi g^{MN} \Big] -
\frac{1}{4} \exp(2 \lambda \varphi) {\cal F}_{MN} {\cal F}^{MN} \},
\end{equation}
where $g = g_{MN} dx^{M} \otimes dx^{N}$ is the metric,
${\cal F} = \frac{1}{2} {\cal F}_{MN}
dx^{M} \wedge dx^{N} = d {\cal A} $ is the strength
of the electromagnetic field and  $\varphi$ is the scalar field
(dilatonic field).  Here $\lambda$ is the dilatonic coupling constant
and $D$ is space-time dimension (in notations of \cite{MS1}
$a = - \lambda \sqrt{D-2}$).

For
\begin{equation} %1.2
\lambda^2 = \frac{1}{8}    , \qquad    D = 10, \qquad  \Lambda = 0
\end{equation}
the action  (1.1) describes a part of the bosonic sector for the $N=1$
ten-dimensional Einstein-Yang-Mills  supergravity that occurs in the
low energy limit of superstring theory \cite{GSW}.
Moreover,
\begin{equation} %1.3
\lambda^2 = \lambda^2_s \equiv \frac{1}{D-2}
\end{equation}
corresponds to the tree-level string effective action in $D$ dimensions.
The non-zero "cosmological term" $\Lambda$ occurs for non-critical
string theories (in this case  $\Lambda$ is proportional to the
central charge deficit).

Another case of interest is
\begin{equation} %1.4
\lambda^2 = \lambda^2_0 \equiv \frac{D -1}{D-2},
\end{equation}
which corresponds to the $D$-dimensional theory obtained
by dimensionally reducing the $(D+1)$-dimensional
Kaluza-Klein theory. In that case the scalar field $\varphi$ is
associated with the size of $(D+1)$-th dimension.

\section{Exact solutions}

The field equations corresponding to the action (1.1)
have the following form and
\begin{eqnarray} %2.1 -2.3
&&R_{MN} -\frac{1}{2} g_{MN} R  =  \kappa^{2} T_{MN}
- \Lambda \exp(- 2\lambda \varphi) g_{MN} , \\
&&\nabla_{M}( \nabla^{M} \varphi) - \frac{1}{2} \lambda
\kappa^{2} \exp(2\lambda \varphi) {\cal F}_{MN} {\cal F}^{MN}  +
2 \lambda \Lambda \exp(- 2 \lambda \varphi) =  0 , \\
&&\nabla_{M} (\exp(2\lambda \varphi) {\cal F}^{MN})  =  0,
\end{eqnarray}
where
\begin{eqnarray}  %2.4
T_{MN} = && \kappa^{-2} (\partial_{M} \varphi \partial_{N}
\varphi  - \frac{1}{2} g_{MN}  \partial_{P} \varphi
\partial^{P} \varphi) \nonumber \\
&& + \exp(2\lambda \varphi) ({\cal F}_{MP} {{\cal F}_{N}}^{P}
- \frac{1}{4} g_{MN} {\cal F}_{PQ} {\cal F}^{PQ}).
\end{eqnarray}

Let us consider the manifold
\begin{equation} %2.5
M = M^{(2+d)} \times M_{int},
\end{equation}
where  $M^{(2+d)}$ is the (2+d)-dimensional (space-time) manifold
and $M_{int}$ is an "internal" space  equipped by the
Ricci-flat metric $g_{int}$.

Our solution to the field equations is defined on the
manifold (2.5) and has the following form
\ba %2.6 -2.8
&&g = - U^{(3-D)2/A}  dt \otimes dt +
      U^{2/A} \Big[ \sum_{a=1}^{1+d}
      dx^a \otimes dx^a +  g_{int} \Big], \\
&&\exp(2\lambda \varphi) = C^2 U^{-2 \alpha_{t}}, \\
&&{\cal A} = {\cal A}_M dx^M  = \frac{\nu dt}{\kappa C U},
\ea
where $C \neq 0$ is constant,
\ba   %2.9 - 2.13
&&A = {A}(\lambda,D) = D-3 + \lambda^2 (D-2), \\
&&\alpha_{t} = -  \lambda^{2}  (D-2)/{A}(\lambda,D), \\
&&\nu^2 =  (D-2)/{A}(\lambda,D), \\
&&U ={U}(t, \vec{x}) = ht + {\Phi}(\vec{x}), \\
&&\bigtriangleup \Phi = \delta^{ab} \partial_a \partial_b
\Phi = 0,
\ea
and
\be   % 2.14
h^2 \frac{(D-2)}{{A}(\lambda,D)}
\Big[ \frac{2(D-2)}{{A}(\lambda,D)}  - 1 \Big] = 2 \Lambda.
\ee
Here $\vec{x} = (x^a)$, $a,b=1, \ldots , 1+d$. It may
be verified by a straightforward calculation that the
equations of motions (2.1)-(2.3) for the considered solution
are satisfied identically  \cite{IM1}.

The solution (2.6)-(2.14) generalizes the Maki-Shiraishi
solution \cite{MS1} (see also \cite{MS2} for $\lambda^2 = 1/2$
and \cite{S1} for $\Lambda = 0$)
to the case of a Ricci-flat internal space. Here we use the
parametrization similar to that of \cite{HH} ($D = 4$).
For $D=d+2$ our notations are related with those from
\cite{MS1} by the following  manner: $A/h = a^2 t_0$,
$(ht)^{a^2/A} = t_{MS}/t_0$.

>From (2.14) we get
\be %2.15
\lambda^2 < \lambda^2_0 \equiv \frac{D-1}{D-2}, \qquad h \neq 0
\ee
for $\Lambda > 0$,
\be %2.16
\lambda^2 > \lambda^2_0,  \qquad h \neq 0
\ee
for $\Lambda < 0$, and
\ba %2.17 -%2.18
&&\lambda^2 = \lambda^2_0, \ \ h \ {\rm is \ arbitrary},  \\
&&{\rm or} \ \lambda \ {\rm is \ arbitrary}, \ \ h = 0
\ea
for $\Lambda = 0$.

Special cases of the above solutions with  $\Lambda > 0$
were also considered by Kastor and Traschen \cite{KT}
($D=4$, $d=2$, $\lambda = 0$) and Horne and Horowitz \cite{HH}
($D= 4$, $\lambda^2 = 1/2$). The solution \cite{KT} generalizes
the well-known Majumdar-Papapetrou solution \cite{MPH}.
For $\Lambda \neq 0$ and
\be  %2.19
\Phi = 1 + \sum_{i=1}^{m} \frac{B_i}{|\vec{x} - \vec{x}_i|^{d-1}}
\ee
the relations (2.6)-(2.8) describe a collection of $m$
multidimensional extremal (charged) dilatonic black holes
(with masses proportional to $B_i$) living in asymptotically
de Sitter or anti-deSitter spaces. Indeed,
as will be shown below, for $\Lambda = 0$
and $m = 1$ in (2.19) we get an $0(d+1)$-symmetric extremal
black hole solution with a Ricci-flat internal space
\cite{BI,BM}. It should be also
noted that the global properties of static spherically
symmetric solutions to the Einstein-Maxwell-dilaton
system in the presence of an arbitrary exponential
dilaton potential were derived in \cite{PW}.

%%%%%%%%%%%%%%%%%%%%%%%%%%%%%%%%%%%%%%%%%%%%%%%%%%%%%%%%%%%%%%%%%%%%

\section{Extremal and non-extremal charged dilatonic black
holes for $\Lambda = 0$.}

Let us consider the case $\Lambda = h = 0$. Here we show that the
solution (2.6)-(2.14) with $\Phi$ from
 (2.19) and $m=1$ is an extremal limit
for the multidimensional charged dilatonic black hole
solution  \cite{BI,BM} (see also \cite{IM}).

\subsection{Non-extremal multidimensional charged dilatonic
black hole}

The solution \cite{BI} with one internal
Ricci-flat space reads
\ba %3.1 -3.3
&&g = - f_{+} f_{-}^{1 + 2\alpha_{t}} dt \otimes dt +
 f_{-}^{ 2\alpha_{r}} \Big[\frac{ dr \otimes dr }{ f_{+} f_{-}  }
 + r^{2} d \Omega^{2}_{d} \Big]
 + f_{-} ^{-2/A}  g_{int},  \\
&& {\cal A} = \Big( \frac{Q}{C(d-1)} r^{1-d} + c \Big) dt, \\
&& \exp(2\lambda \varphi) = C^2 f_{-}^{2\alpha_{t}}.
\ea
Here
\ba %3.4 -3.5
&&f_{\pm} = {f_{\pm}}(r) =  1 - \frac{B_{\pm}}{r^{d-1}},  \\
&&\alpha_{r} = \frac{1}{d-1} -  \frac{1}{{A}(\lambda,D)},
\ea
$g_{int}$   is Ricci-flat  internal space metric, $d \Omega^{2}_{d}$ is
canonical metric on $S^d$
("spherical angle"), parameters $A = {A}(\lambda,D)$ and $\alpha_{t}$ are
defined in (2.9) and (2.10) respectively,
$c$ and $C \neq 0$ are constants and the parameters
\be %3.6
B_{+} > B_{-} > 0, \qquad Q \neq 0
\end{equation}
satisfy the relation
\be %3.7
B_{+} B_{-} = \frac{\kappa^{2} Q^{2}
{A}(\lambda,D)} {(d-1)^{2}(D-2)}.
\ee
(Here we have slightly generalized \cite{BI} by introducing the
constant $C \neq 0$.)

The $(2+d)$-dimensional section of the metric (3.1) has
a horizon at  $r^{d-1} = B_{+}$. For  $r^{d-1} = B_{-} < B_{+}$
a horizon is absent since
\be
\alpha_{t} - \alpha_{r} < 0.
\ee

We note that $\alpha_{t} \leq 0$, $\alpha_{r} \geq 0$ and
$\alpha_{r} = 0  \Leftrightarrow (\lambda = 0, \ D = d + 2)$.
The horizon at $r^{d-1} = B_{-} < B_{+}$
takes place only for $\lambda = 0$ and $D = d + 2$. In this case the
internal space is absent and we are led to the Myers-Perry
$O(d+1)$-symmetric charged black hole solution \cite{MP}.

The considered solution generalizes the solutions
from \cite{GM,GHS} with $D= 2 + d$.

The mass of the black hole is defined by the relation \cite{BI}
\begin{equation}  %3.9
2 G M =  B_{+} + B_{-} \beta,
\end{equation}
where
\begin{equation}  %3.10
\beta = 1 + 2 \alpha_{t} = \frac{D-3 - \lambda^{2}(D-2)}
{D-3 + \lambda^{2}(D-2)}
\end{equation}
and $G  =  S_{D} \kappa^{2}$  is the effective gravitational
constant ($S_{D}$ is defined in \cite{MP}).

{\bf Hawking temperature}. A standard calculation based on the
absence of conic singularity as $r^{d-1} \to B_{+}$  in the Euclidean-
rotated metric (3.1) ($t = -i \tau$, $0 \leq \tau \leq T_H^{-1}$)
gives us the following relation
for the Hawking temperature
\be %3.11
T_H = \frac{(d-1)}{4 \pi r_{+}}
\Big(1 - \frac{B_{-}}{B_{+}} \Big)^{\alpha},
\ee
where $r_{+} = (B_{+})^{1/(d-1)}$ and
\be  %3.12
\alpha  = 1 + \alpha_{t} - \alpha_{r} < 1.
\ee
For the string case (1.3) $\alpha = \alpha_s = (d-2)/(d-1) \geq 0$.
>From (3.12) we get the inequality
\be %3.13
T_H  > T_{MP} = \frac{(d-1)}{4 \pi r_{+}}
\Big(1 - \frac{B_{-}}{B_{+}} \Big)
\ee
where $T_{MP}$ is the Hawking temperature for the Myers-Perry
charged black hole \cite{MP} ($\lambda = 0$, $D = d + 2$).
For $D=4$ ($d=2$) relation (3.11) agrees with the corresponding
relation from \cite{S2}.

\subsection{Extremal charged black hole}

Now we consider the solution (3.1)-(3.5), (3.6) for
\begin{equation} %3.14
B_{+} = B_{-} =
\frac{\kappa |Q| \sqrt{{A}(\lambda,D)}}{(d-1) \sqrt{(D-2)}} = B.
\end{equation}

In this case the mass (3.9) is minimal \cite{BI}
\begin{equation} %3.15
M_{c} =  \frac{\kappa |Q| (D-3)}{G(d-1) \sqrt{{A}(\lambda,D)(D-2)}}
\end{equation}
and the relations (3.1)-(3.5) describe an extremal charged
dilatonic black hole. Introducing a new radial variable $R$ by
the relation
\be %3.16
r^{d-1} - B = R^{d-1}
\ee
and denoting
\be %3.17
U = 1 + \frac{B}{R^{d-1}} = f_{\pm}
\ee
we get a special case of the solution from Sec. 2 with
$\Lambda = h = 0$ and $m = 1$ in (2.19) ($R = |\vec{x} - \vec{x}_1|$,
$B = B_1$). Here we put $c = - Q/(B(d-1))$ in (3.2).

In the extremal case a horizon for  $r^{d-1} = B$
takes place if (and only if)
\be  %3.18
\alpha  = {\alpha}(\lambda,d,D) \geq 0
\ee
(see (3.12)). This is equivalent to the following
restriction on the dilatonic coupling parameter
\be %3.19
\lambda^2 \leq d - 2 + \frac{1}{(D-2)} \equiv \lambda^2_c.
\ee

The relation (3.19) is satisfied for the string case (1.3).
For $d=2$ we have $\lambda^2_c = \lambda^2_s$.

For $\lambda^2 = \lambda^2_c$  we get $\alpha = 0$
and
\be %3.20
T_H = T_c \equiv \frac{d-1}{4\pi r_{+}}.
\ee

Due to relation (3.11) the Hawking temperature has the following limit as
$B_{-} \to B_{+}$, i.e. in the extreme black hole limit
\ba %3.21 -22
T_H \to &&0, \qquad  {\rm for} \ \lambda^2 < \lambda^2_c, \\
        &&T_c, \qquad  {\rm for} \ \lambda^2 = \lambda^2_c.
\ea

For $\lambda^2 > \lambda^2_c$ we get $T_H \to + \infty$
as $B_{-} \to B_{+}$. In this case the horizon at
$r^{d-1} = B_{-} = B_{+}$ is absent.

Remark. It may be shown that the metric (3.1) is
singular as $r \to r_{-} = B_{-}^{1/(d-1)}$  for all $B_{+} \geq B_{-} >
0$, $\lambda$ and $D > d+2$ \cite{IM1}. For $\lambda = 0$ and $D = d+2$
(i.e. in the Myers-Perry case \cite{MP}) the singularity is absent.

\section{Special cases}

\subsection{Infinite-dimensional case}

Here we consider the interesting special case, when
$N_{int} = {\rm dim} M_{int} \to + \infty$
(or $D \to + \infty$ and $d$ is fixed).
In this limit the metric (2.6) reads
\be %4.1
g = - U^{- 2/(1 + \lambda^2)}  dt \otimes dt +
      \sum_{a=1}^{1+d}  dx^a \otimes dx^a +  g_{int},
\ee
and the relations (2.10), (2.11) and (2.14) take the following
form
\ba   %4.2 - 4.3
&&\alpha_{t} = - \lambda^{2}/(1 + \lambda^{2}) ,  \qquad
\nu^2 =  1/(1 + \lambda^{2}), \\
&& h^2 \frac{1 - \lambda^{2}}{(1 + \lambda^{2})^2} = 2 \Lambda.
\ea
Thus, the spatial section of the metric (4.1)  is flat.
For string (1.3), "Kaluza-Klein" (1.4) and critical  (3.19) values of
the coupling constants we have respectively
\be   %4.4
\lambda^{2}_s = 0, \qquad
\lambda^{2}_0 = 1, \qquad \lambda^{2}_c = d-2.
\ee
For the parameter $\alpha$  in (3.12) we have
\be   %4.5
\alpha =  \frac{1}{1 + \lambda^{2}} - \frac{1}{d -1}.
\ee

\subsection{Stringy case}

For  $\lambda^{2} = \lambda^{2}_s = (D-2)^{-1}$  we
obtain instead of (4.1)-(4.3) the following relations
\be %4.6
g = - U^{(3 - D)2/(D-2)}  dt \otimes dt +
U^{2/(D-2)}
\Big[ \sum_{a=1}^{1+d}  dx^a \otimes dx^a +  g_{int} \Big],
\ee
\ba   %4.7 - 4.8
&&\alpha_{t} = - (D-2)^{-1} ,  \qquad
\nu^2 =  1, \\
&& h^2  = 2 \Lambda.
\ea

For the Hawking temperature (3.11) we get in the considered case
\be %4.9
T_H = T_{H,s} = \frac{(d-1)}{4 \pi r_{+}}
\Big(1 - \frac{B_{-}}{B_{+}} \Big)^{\frac{d-2}{d-1}}.
\ee

Thus, in the string case (1.3)
the Hawking temperature  (4.9)
does not depend upon the total
dimension $D$ (or equivalently upon the internal space dimension
$N_{int}$). Moreover, it does not depend upon the
Ricci-flat internal  space $(M_{int}, g_{int})$.
For example, one may consider for $D=10$ different internal Calabi-Yau
spaces \cite{GSW} with the same result for $T_H$.
For $d = 2$ (4.9) coincides with the corresponding
formula for the Schwarzschild black hole.

\begin{center}
{\bf Acknowledgments}
\end{center}

The authors are grateful to K.A. Bronnikov and M.L. Fil'chenkov,
for useful discussions.  This work was supported  in part by the
Russian Ministry of Science and Russian Fund of Basic Sciences.

\pagebreak

\end{document}